\documentclass[12pt,draftclsnofoot,onecolumn]{IEEEtran}

\usepackage{comment,url}
\usepackage[dvips]{graphicx}
\usepackage{caption,subfigure,color}
\usepackage{float}
\graphicspath{{./figure/}}
\usepackage{setspace}
\usepackage{amsmath}
\usepackage{amssymb}
\usepackage{algorithm}
\usepackage{algorithmic}
\usepackage{cite}
\usepackage{array}
\usepackage{mdwmath}
\usepackage{mdwtab}

\allowdisplaybreaks[4]
\long\def\symbolfootnote[#1]#2{\begingroup%
\def\thefootnote{\fnsymbol{footnote}}\footnote[#1]{#2}\endgroup}

\begin{document}
%
\title{Random Distances Associated with Arbitrary Polygons: An Algorithmic Approach\\between Two Random Points}

\author{Fei Tong and Jianping Pan\\
University of Victoria, Victoria, BC, Canada}

\maketitle
\begin{abstract}
This report presents a new, algorithmic approach to the distributions of the distance between two points distributed uniformly at random in various polygons, based on the extended Kinematic Measure (KM) from integral geometry. We first obtain such random Point Distance Distributions (PDDs) associated with arbitrary triangles (i.e., triangle-PDDs), including the PDD within a triangle, and that between two triangles sharing either a common side or a common vertex. For each case, we provide an algorithmic procedure showing the mathematical derivation process, based on which either the closed-form expressions or the algorithmic results can be obtained. The obtained triangle-PDDs can be utilized for modeling and analyzing the wireless communication networks associated with triangle geometries, such as sensor networks with triangle-shaped clusters and triangle-shaped cellular systems with highly directional antennas. Furthermore, based on the obtained triangle-PDDs, we then show how to obtain the PDDs associated with arbitrary polygons through the decomposition and recursion approach, since any polygons can be triangulated, and any geometry shapes can be approximated by polygons with a needed precision. Finally, we give the PDDs associated with ring geometries. The results shown in this report can enrich and expand the theory and application of the probabilistic distance models for the analysis of wireless communication networks.
\end{abstract}


\section{PDD within a Triangle}\label{subsec:within-triangle}
\begin{figure}[H]
\centering
\includegraphics[width=6in]{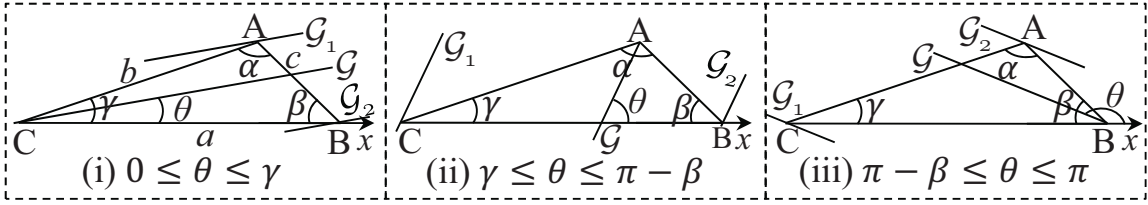}
\vspace{-.em}
\caption{KM over an arbitrary triangle.}
\label{fig:triangle-theta}
\vspace{-2.em}
\end{figure}
$\triangle{}ABC$ is an arbitrary triangle with side lengths $|CB|=a$, $|AC|=b$, and $|AB|=c$, internal angles $\angle A=\alpha$, $\angle B=\beta$, and $\angle C=\gamma$, and area $||\triangle ABC||=S$. Without loss of generality (WLOG), $a\geq b\geq c$, and let side $CB$ be on $x$-axis, as shown in \figurename~\ref{fig:triangle-theta}. For simplicity, we can have $a=1$ and other edges normalized correspondingly. The Probability Density Function (PDF) of the PDD, denoted as $f_D(d)$, can be scaled to any size of triangles with $a=s$ by
\begin{equation}\label{eq:pdf_scale}
  f_{sD}=\frac{1}{s}f_D(\frac{d}{s})~,
\end{equation}
where $f_{sD}$ is the corresponding PDF of the PDD within the triangle with $a=s$. Such a scaling is applicable to any polygons. When calculating the length of the chord produced by a line intersecting with the triangle with orientation $\theta$ with regard to $x$-axis, there are three cases in terms of the range of $\theta$: (i) $0\leq\theta\leq\gamma$, (ii) $\gamma\leq\theta\leq\pi-\beta$, (iii) $\pi-\beta\leq\theta\leq\pi$. We then design a systematic algorithmic procedure for the numerical integration of the intended PDD, based on which the corresponding closed-form expression is also derived, as shown below in detail.

\begin{figure}[!t]
\renewcommand{\algorithmicrequire}{\textbf{Input:}}
\renewcommand\algorithmicensure {\textbf{Output:}}
 \rule[-20pt]{1\textwidth}{0.12em}
 \begin{algorithmic}[1]
 \REQUIRE Parameters with regard to $\triangle ABC$ in \figurename~\ref{fig:triangle-theta}.
 \ENSURE PDF: $f_D(d)$.
 \STATE $f_D(d)=0$;
 \FOR{($\theta=0$; $\theta\leq\pi$; $\theta=\theta+\delta\theta$)}
    \IF{$0\leq\theta\leq\gamma$}\label{alg1:1}
        \STATE $p_1=b\sin(\gamma-\theta)$; $p_2=a\sin(\theta)$; $base=\frac{b\sin(\alpha)}{\sin(\theta+\beta)}$;
    \ELSIF{$\gamma\leq\theta\leq\pi-\beta$}
        \STATE $p_1=b\sin(\theta-\gamma)$; $p_2=c\sin(\theta+\beta)$; $base=\frac{c\sin(\beta)}{\sin(\theta)}$;
    \ELSIF{$\pi-\beta\leq\theta\leq\pi$}
        \STATE $p_1=a\sin(\theta)$; $p_2=-c\sin(\theta+\beta)$; $base=\frac{c\sin(\alpha)}{\sin(\theta-\gamma)}$;
    \ENDIF\label{alg1:2}
    \STATE $p_m=p_1+p_2$; $/*$ width $*/$
    \FOR{($p=0$; $p\leq p_m$; $p=p+\delta p$)}
        \IF{$p\leq p_1$}
            \STATE $l=\frac{p\cdot base}{p_1}$;\label{alg1:3}
        \ELSE
            \STATE $l=\frac{(p_m-p)\cdot base}{p_2}$;
        \ENDIF
        \IF{$l\geq d$}
           	\STATE $f_D(d)=f_D(d)+\frac{2d(l-d)}{S^2}$; \label{alg1:4}
        \ENDIF
    \ENDFOR
 \ENDFOR
 \end{algorithmic}
 \vspace{-1em}
 \rule[0pt]{1\textwidth}{0.05em}
  \vspace{-1.em}
 \caption{Algorithmic procedure for obtaining the PDD within an arbitrary triangle.}\label{alg:triangle}
 \vspace{-2.em}
\end{figure}%
Specifically, let $\theta$ increase from $0$ to $\pi$ with a fixed small step of $\delta\theta$ (e.g., $\delta\theta=\frac{\pi}{180}$). For each $\theta$ with a set of lines intersecting with the triangle, $\mathcal{G}$ is the line which produces the longest chord of length $base$. The distance between the two tangents parallel with $\mathcal{G}$, i.e., the support lines $\mathcal{G}_1$ and $\mathcal{G}_2$ which completely encompass the whole triangle, is $p_m$. The distance between $\mathcal{G}_1$ and $\mathcal{G}$ and that between $\mathcal{G}_2$ and $\mathcal{G}$ are $p_1$ and $p_2$, respectively. Obviously, $p_m=p_1+p_2$. With $p$ increasing from 0 to $p_m$ with a fixed small step $\delta{}p$ (e.g., $\delta{}p=\frac{1}{1,000}$), we obtain $\frac{p_m}{\delta p}$ chords. For each chord of length $l$ calculated based on trigonometry, we obtain $f_{\mathcal{G}}(d)$, based on which the PDF of the PDD can be calculated. The derivation is summarized in \figurename~\ref{alg:triangle}, providing the regularity to help obtain the PDF symbolically as $\delta\theta$ and $\delta p$ go to $0$.

We investigate the case (i) where $0\leq\theta\leq\gamma$ first, and show below how to obtain the corresponding closed-form expression based on the algorithmic procedure shown in \figurename~\ref{alg:triangle}. Specifically, for $0\leq p\leq p_1$, a chord is determined by ($\theta,p$) with length of $l=\frac{p\cdot base}{p_1}$ as shown in line \ref{alg1:3} of \figurename~\ref{alg:triangle}. From $l\geq d$ (line \ref{alg1:4} of \figurename~\ref{alg:triangle}), the integration range of $p$ is $[\frac{d\cdot p_1}{base},p_1]$, and $\theta\leq\theta^{\rm i}_1=\arcsin\left(\frac{b\sin(\alpha)}{d}\right)-\beta$ or $\theta\geq\theta^{\rm i}_2=\pi-\arcsin\left(\frac{b\sin(\alpha)}{d}\right)-\beta$. Meanwhile, for $p_1\leq p\leq p_m$, the length of the determined chord is $l=\frac{(p_m-p)\cdot base}{p_2}$. Similarly, from $l\geq d$, we have $p\in[p_1,p_m-\frac{d\cdot p_2}{base}]$, and $\theta\leq\theta^{\rm i}_1$ or $\theta\geq\theta^{\rm i}_2$. Therefore,
\begin{equation}\label{eq:f_I}
f_D^{\rm i}(d)=
\left\{
\begin{array}{ll}
f^{\rm i}_1 & ~~~ \gamma\leq\frac{\pi}{2}-\beta\\
f^{\rm i}_{21}+f^{\rm i}_{22} & ~~~ \mathrm{otherwise}
\end{array}
\right.,
\end{equation}%
where
\begin{align}
&f^{\rm i}_1=
\left\{
  \begin{array}{ll}
    H^{\rm i}_1\left(0,\theta^{\rm i}_1\right)+H^{\rm i}_2\left(0,\theta^{\rm i}_1\right) & 0\leq\theta^{\rm i}_1\leq\gamma\\
    H^{\rm i}_1(0,\gamma)+H^{\rm i}_2(0,\gamma) & \theta^{\rm i}_1>\gamma\\
  0 & \mathrm{otherwise}\\
  \end{array}
\right.,\nonumber\\
&f^{\rm i}_{21}=
\left\{
  \begin{array}{ll}
    H^{\rm i}_1\left(0,\theta^{\rm i}_1\right)+H^{\rm i}_2\left(0,\theta^{\rm i}_1\right) & 0\leq\theta^{\rm i}_1\leq \frac{\pi}{2}-\beta\\
    0 & \mathrm{otherwise}
  \end{array}
\right.,\nonumber\\
&f^{\rm i}_{22}=
\left\{
  \begin{array}{ll}
    H^{\rm i}_1\left(\theta^{\rm i}_2,\gamma\right)+H^{\rm i}_2\left(\theta^{\rm i}_2,\gamma\right) & \theta^{\rm i}_2\leq\gamma\\
    0 & \mathrm{otherwise}
  \end{array}
\right.,\nonumber\\
&\begin{array}{rl}\label{H1-x-y}
H^{\rm i}_1(\mathcal{X},\mathcal{Y})=&\frac{1}{S^2}\int_\mathcal{X}^\mathcal{Y}\int_{\frac{d\cdot p_1}{base}}^{p_1}2d\left(l_1-d\right )~\mathrm{d}p\mathrm{d}\theta~,
\end{array}\\
&\begin{array}{rl}\label{H2-x-y}
H^{\rm i}_2(\mathcal{X},\mathcal{Y})=&\frac{1}{S^2}\int_\mathcal{X}^\mathcal{Y}\int_{p_1}^{p_m-\frac{d\cdot p_2}{base}}2d\left(l_2-d\right)~\mathrm{d}p\mathrm{d}\theta~.
\end{array}
\end{align}

Similarly, for case (ii), we have
\begin{equation}\label{eq:f_II}
f_D^{\rm ii}(d)=f^{\rm ii}_1+f^{\rm ii}_2~,
\end{equation}%
where
\begin{align*}
  &f^{\rm ii}_1=
    \left\{
      \begin{array}{ll}
        H^{\rm ii}_1(\gamma,\theta^{\rm ii}_1)+H^{\rm ii}_2(\gamma,\theta^{\rm ii}_1) & \gamma\leq\theta^{\rm ii}_1\leq\frac{\pi}{2}\\
        0 & \mathrm{otherwise}
      \end{array}
    \right.,\\
  &f^{\rm ii}_2=
    \left\{
    	\begin{array}{ll}
    		H^{\rm ii}_1(\theta^{\rm ii}_2,\pi-\beta)+H^{\rm ii}_2(\theta^{\rm ii}_2,\pi-\beta) & \theta^{\rm ii}_2\leq\pi-\beta\\
    		0 & \mathrm{otherwise}
    	\end{array}
    \right.,\\
    &\begin{array}{l}
    \theta^{\rm ii}_1=\arcsin\left(\frac{c\sin(\beta)}{d}\right)~,\\
    \theta^{\rm ii}_2=\pi-\arcsin\left(\frac{c\sin(\beta)}{d}\right)~,
    \end{array}
\end{align*}%
and for case (iii),
\begin{equation}\label{eq:f_III}
f_D^{\rm iii}(d)=
\left\{
\begin{array}{ll}
f^{\rm iii}_1 & ~~~ \beta\leq\frac{\pi}{2}-\gamma\\
f^{\rm iii}_{21}+f^{\rm iii}_{22} & ~~~ \mathrm{otherwise}
\end{array}
\right.,
\end{equation}%
where
\begin{align*}
&f^{\rm iii}_1=
\left\{
  \begin{array}{ll}
    H^{\rm iii}_1(\pi-\beta,\pi)+H^{\rm iii}_2(\pi-\beta,\pi) & \theta^{\rm iii}_1<\pi-\beta\\
    H^{\rm iii}_1(\theta^{\rm iii}_1,\pi)+H^{\rm iii}_2(\theta^{\rm iii}_1,\pi) & \pi-\beta\leq\theta^{\rm iii}_1\leq\pi\\
  0 & \mathrm{otherwise}\\
  \end{array}
\right.,\nonumber\\
&f^{\rm iii}_{21}=
\left\{
  \begin{array}{ll}
    H^{\rm iii}_1(\pi-\beta,\theta^{\rm iii}_2)+H^{\rm iii}_2(\pi-\beta,\theta^{\rm iii}_2) & \pi-\beta\leq\theta^{\rm iii}_2\leq \frac{\pi}{2}+\gamma\\
    0 & \mathrm{otherwise}
  \end{array}
\right.,\\
&f^{\rm iii}_{22}=
\left\{
  \begin{array}{ll}
    H^{\rm iii}_1(\theta^{\rm iii}_1,\pi)+H^{\rm iii}_2(\theta^{\rm iii}_1,\pi) & \theta^{\rm iii}_1\leq\pi\\
    0 & \mathrm{otherwise}
  \end{array}
\right.,\\
&\begin{array}{l}
\theta^{\rm iii}_1=\pi-\arcsin(\frac{c\sin(\alpha)}{d})+\gamma,\\
\theta^{\rm iii}_2=\arcsin(\frac{c\sin(\alpha)}{d})+\gamma.
\end{array}
\end{align*}%
Finally, the PDF of the PDD within an arbitrary triangle is
\begin{equation}\label{eq:triangle-gdd}
\begin{array}{l}
   f_D(d)=f_D^{\rm i}(d)+f_D^{\rm ii}(d)+f_D^{\rm iii}(d)~.
\end{array}
\end{equation}
Similar to $H^{\rm i}_1$ and $H^{\rm i}_2$, $H^{\rm ii}_1$ and $H^{\rm iii}_1$ can be calculated using (\ref{H1-x-y}), and $H^{\rm ii}_2$ and $H^{\rm iii}_2$ can be calculated using (\ref{H2-x-y}), but with different $p_1$, $p_2$ and $base$ for different cases (see line \ref{alg1:1}--\ref{alg1:2} of \figurename~\ref{alg:triangle}). $H^{\rm i}_1$, $H^{\rm i}_2$, $H^{\rm ii}_1$, $H^{\rm ii}_2$, $H^{\rm iii}_1$, and $H^{\rm iii}_2$ in (\ref{eq:triangle-gdd}) are
\begin{equation*}
 \renewcommand{\arraystretch}{1.5}
 \left\{\begin{array}{l}
    H^{\rm i}_1(\mathcal{X},\mathcal{Y})=h^{\rm i}_1(\mathcal{Y})-h^{\rm i}_1(\mathcal{X}),
    H^{\rm i}_2(\mathcal{X},\mathcal{Y})=h^{\rm i}_2(\mathcal{Y})-h^{\rm i}_2(\mathcal{X}),\\
    H^{\rm ii}_1(\mathcal{X},\mathcal{Y})=h^{\rm ii}_1(\mathcal{Y})-h^{\rm ii}_1(\mathcal{X}),
    H^{\rm ii}_2(\mathcal{X},\mathcal{Y})=h^{\rm ii}_2(\mathcal{Y})-h^{\rm ii}_2(\mathcal{X}),\\
    H^{\rm iii}_1(\mathcal{X},\mathcal{Y})=h^{\rm iii}_1(\mathcal{Y})-h^{\rm iii}_1(\mathcal{X}),
    H^{\rm iii}_2(\mathcal{X},\mathcal{Y})=h^{\rm iii}_2(\mathcal{Y})-h^{\rm iii}_2(\mathcal{X})~,
  \end{array}\right.\\
\end{equation*}%
where
$$
{\allowdisplaybreaks
 \renewcommand{\arraystretch}{1.15}
\begin{array}{rl}
  h^{\rm i}_1(\theta)&=\frac{d}{2\sin ( \alpha ) }( \frac{{d}^{2}}{2}\sin ( \beta-\gamma+2\,\theta ) -d (4\,b\sin ( \alpha )\cos ( \gamma-\theta )+\\
    &d\theta\cos( \beta+\gamma )  ) + \frac{{b}^{2}}{2}\ln  ( -{\frac {\sin ( \beta+\theta ) }{\cos ( \gamma-\theta ) }} ) ( 2\,\sin ( \beta+\gamma )-\\
    &\sin ( 2\,\alpha+\beta+\gamma ) +\sin ( 2\,\alpha-\beta-\gamma )  ) +\sin ^{2}( \alpha )(2\,{b}^{2} \\
    &( \gamma-\theta )  \cos ( \beta+\gamma )-{b}^{2}\ln  (  \tan^{2} ( \gamma-\theta )  +1 ) \sin ( \beta+\gamma ))   )~,\\
    h^{\rm i}_2(\theta)&={\frac {ad}{b\sin ( \alpha ) }} ( \frac{{d}^{2}\theta}{2}\cos ( \beta ) -\frac{{d}^{2}}{4}\sin( \beta+2\,\theta )+{b}^{2}\theta\cos ( \beta )  \sin ^{2} ( \alpha )\\
    {}&+2\,bd\sin ( \alpha ) \cos ( \theta ) -{b}^{2}\ln  ( \sin ( \beta+ \theta )  ) \sin ( \beta )  \sin^2 ( \alpha ) )~,\\
    h^{\rm ii}_1(\theta)&=\frac {bd}{4c\sin ( \beta ) } ( {d}^{2}\sin ( \gamma-2\,\theta ) +2\,{d}^{2}\theta\,\cos ( \gamma )-4\,{c}^{2}\sin^{2} ( \beta ) \\
    {}&(\ln  ( \sin (\theta )  )  \sin ( \gamma ) -\theta\,\cos ( \gamma ) ) +8\,cd\sin ( \beta ) \cos ( \gamma-\theta )  ) ~,\\
    h^{\rm ii}_2(\theta)&=\frac {d}{4\sin ( \beta ) } ( 2\,{d}^{2}\theta\,\cos ( \beta ) -{d}^{2}\sin ( \beta+2\,\theta)+4\,{c}^{2}\sin^2 ( \beta )\\
    {}&(\ln  ( \sin (\theta )  )   \sin ( \beta )+\theta\,\cos ( \beta ))+8\,cd\cos ( \beta+\theta ) \sin ( \beta ) ) ~,\\
    h^{\rm iii}_1(\theta)&= \frac {ad}{4c\sin ( \alpha ) } ( {d}^{2}\sin ( \beta-2\,\theta ) +2\,{d}^{2}\theta\,\cos ( \beta )+8\,cd\sin ( \alpha ) \\
    {}& \cos ( \theta )+4\,{c}^{2}\sin^{2} ( \alpha )(\theta\,\cos ( \beta )+\sin ( \beta ) \ln  ( -\sin ( \beta-\theta )  ) )   ) ~,\\
    h^{\rm iii}_2(\theta)&=\frac {2d}{\sin ( \alpha ) } ( \frac{{d}^{2}}{8}\sin ( \beta-\gamma+2\,\theta )-\\
    &\frac{\theta}{4}\cos ( \beta+\gamma )( 2\,{c}^{2} \sin^2 ( \alpha ) +{d}^{2} ) - cd\cos ( \beta+\theta ) \sin ( \alpha )-\\
    &\frac{{c}^{2}}{2}\ln  ( \sin ( \gamma-\theta )  ) \sin ( \beta+\gamma )   \sin^2 ( \alpha )  ) ~.
\end{array}}
$$%
Therefore, the closed-form expression of the PDF of PDD within an arbitrary triangle, i.e., \eqref{eq:triangle-gdd}, has been obtained.

\textcolor{black}{Although the obtained closed-form expression looks tedious, note that, for the network performance analysis, we do not use the symbolic expression of \eqref{eq:triangle-gdd} directly but the numerical PDF result calculated promptly by \eqref{eq:triangle-gdd} providing the necessary parameters (e.g., two edges and one angle, or two angles and one edge) of an arbitrary triangle. Simulations can also be utilized for obtaining PDDs. However, conducting simulations for each specific triangle is very time-consuming, and requires a large number of runs to obtain statistically significant results. Moreover, by simulations, only the empirical Cumulative Distribution Function (CDF) of nodal distances can be obtained, while the accurate PDF is also indispensable to the modeling and analysis of wireless communication networks.}

The obtained results are verified in comparison with simulation and the approach in \cite{Fei'arxiv'13} based on Chord Length Distribution (CLD). The simulation is conducted in Matlab as below (the following simulations are all conducted in a similar way):
\begin{description}
  \item[(1)] Generate a point uniformly at random within a triangle.
  \item[(2)] Generate another point uniformly at random within the triangle.
  \item[(3)] Compute the Euclidean distance between these two points and append the obtained distance to a matrix.
  \item[(4)] Repeat steps (1)--(3) $50,000$ times (the more repeats, the more accurate the result). Then use the Matlab function ``ecdf'' with the matrix as its parameter, we can obtain the empirical CDF of the PDD within the triangle.
\end{description}
\begin{figure}[!t]
\centering
\includegraphics[width=0.6\textwidth]{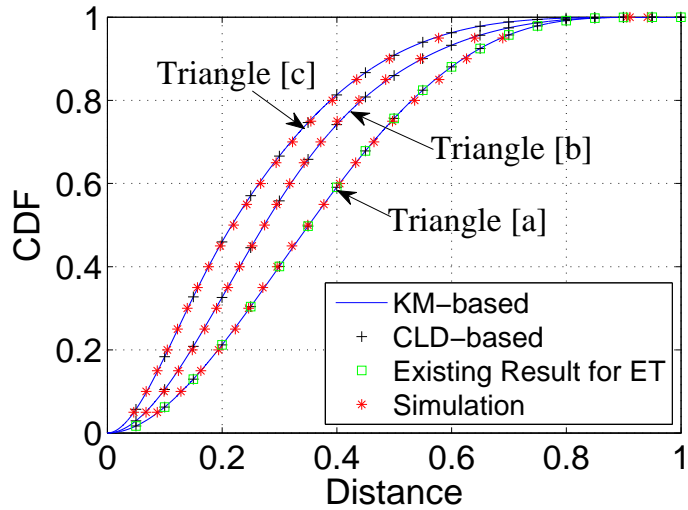}
\vspace{-0em}
\caption{PDDs within arbitrary triangles.}
\label{fig:pdd-1-triangle}
\vspace{-2.em}
\end{figure}%
For simplicity, only numerical CDFs obtained by integrating the corresponding PDFs are shown here and hereafter. Three triangles are selected: {{[a]} $\left(\frac{60\pi}{180},\frac{60\pi}{180},\frac{60\pi}{180}\right)$, {[b]} $\left(\frac{80\pi}{180},\frac{70\pi}{180},\frac{30\pi}{180}\right)$, and {[c]} $\left(\frac{130\pi}{180},\frac{30\pi}{180},\frac{20\pi}{180}\right)$}, all of which have the longest side length of 1, which can be scaled to any nonzero size as introduced in \eqref{eq:pdf_scale}. For the first triangle, i.e., an equilateral triangle (ET), the result is also compared with that obtained in~\cite{Zhuang'arxiv'12}, which is only applicable for ET. Figure~\ref{fig:pdd-1-triangle} shows a close match between the obtained results based on KM and the simulation results.
\section{PDD between Two Triangles}\label{subsec:bet-2-triangles}
\begin{figure}[!t]
\centering  
\subfigure[convex quadrangle]{\includegraphics[width=0.4\textwidth]{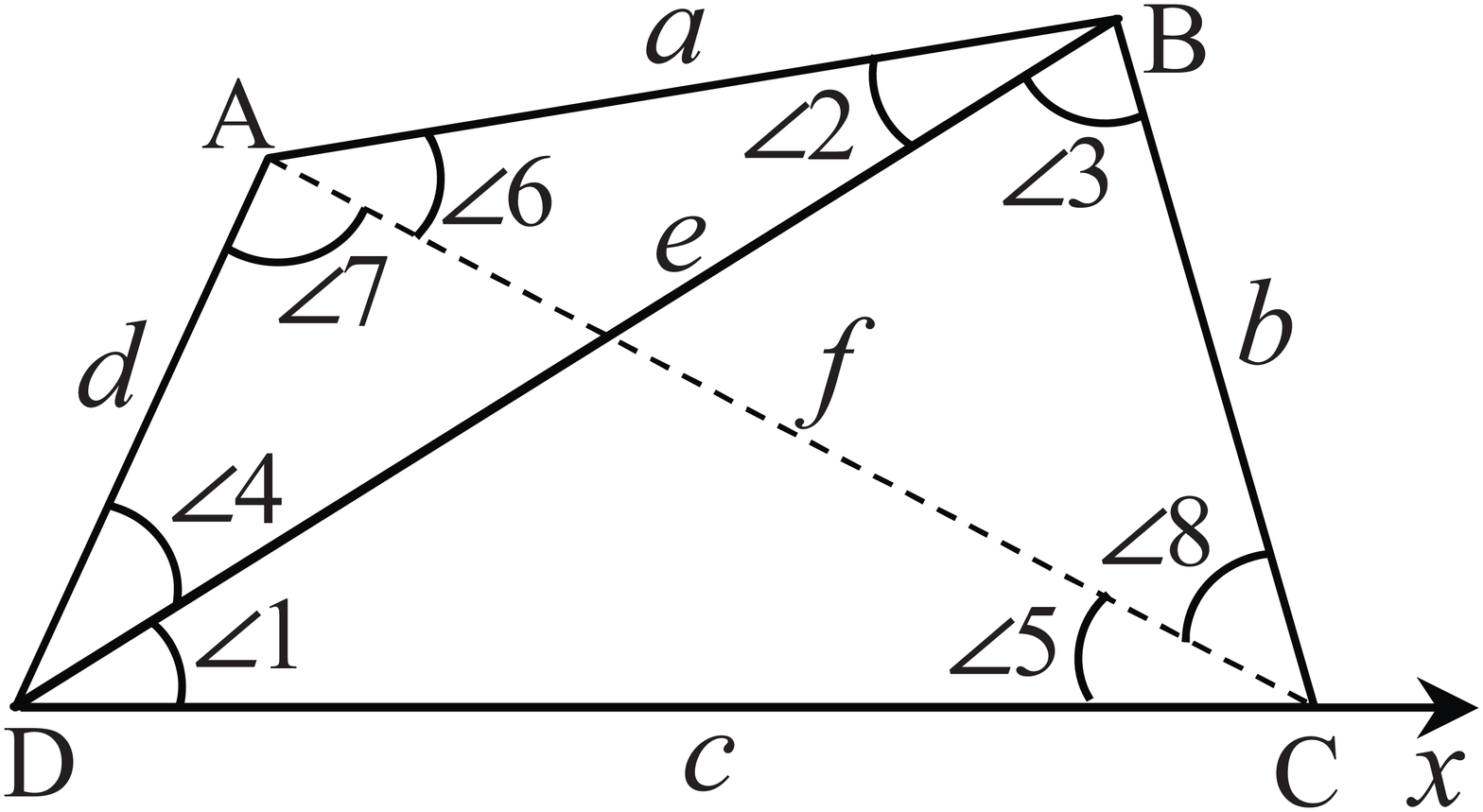}\label{fig:2-triangles-convex}}
\hspace{1.5em}
\subfigure[concave quadrangle]{\includegraphics[width=0.4\textwidth]{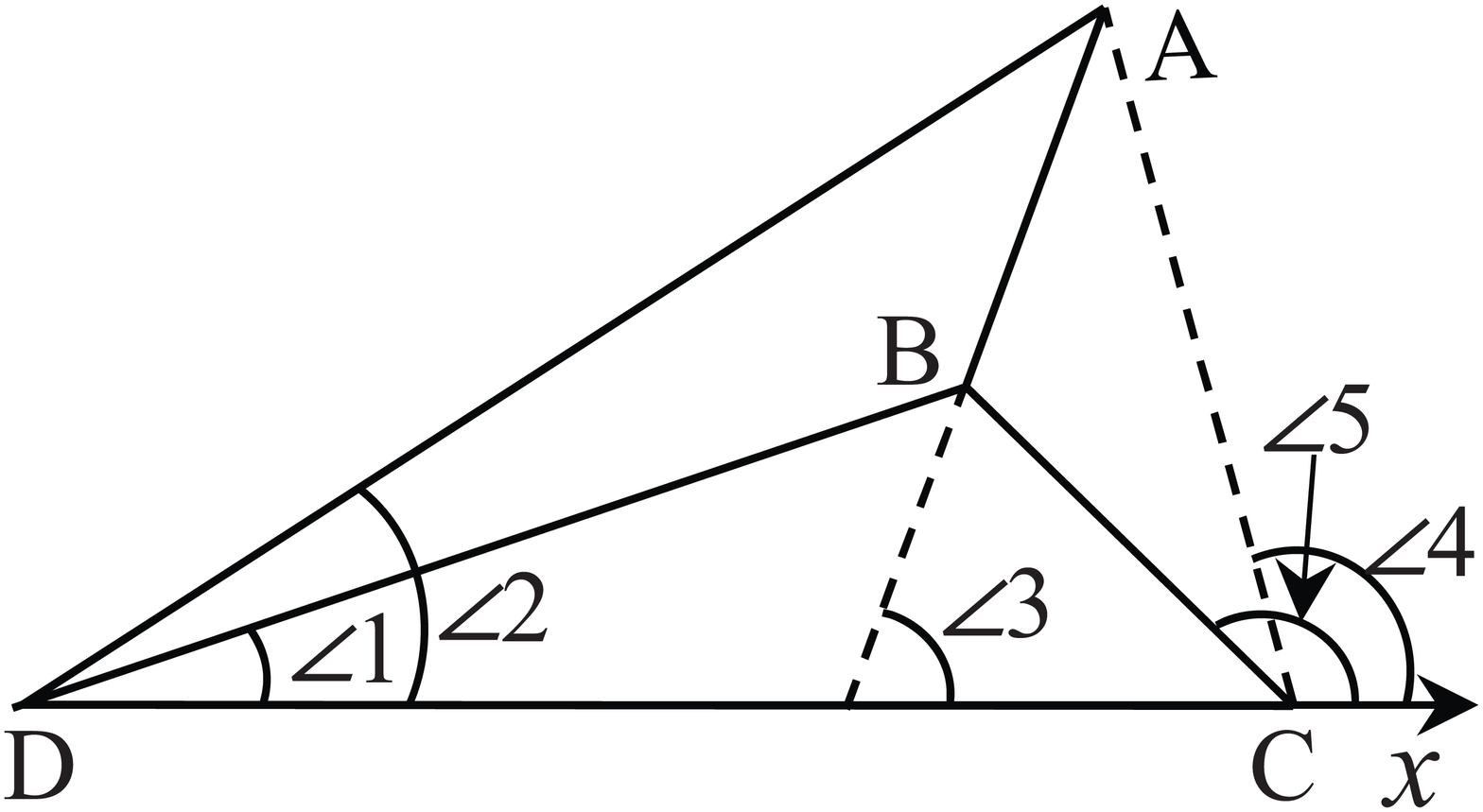}\label{fig:2-triangles-concave}}
\vspace{-.5em}
\caption{Two triangles sharing a common side form a quadrangle ($||\triangle ABD||=S_1$, and $||\triangle BCD||=S_2$).}
\vspace{-2.em}
\end{figure}

Our approach can also be utilized to obtain the PDD between two disjoint geometries, with which the PDD-based performance metrics associated with two clusters in ad-hoc networks or two cells in cellular systems can be quantified. In this section, we show how to obtain the PDDs between two triangles sharing either a common side or a vertex. For the former, the two triangles can form either a convex or concave quadrangle, as shown in \figurename~\ref{fig:2-triangles-convex} and~\ref{fig:2-triangles-concave}, respectively. For simplicity, hereafter, only the algorithmic procedure showing the derivation process is provided, based on which the closed-form expressions can be derived following the same method as shown in Section~\ref{subsec:within-triangle}.
\begin{figure}[!t]
\centering
\includegraphics[width=6in]{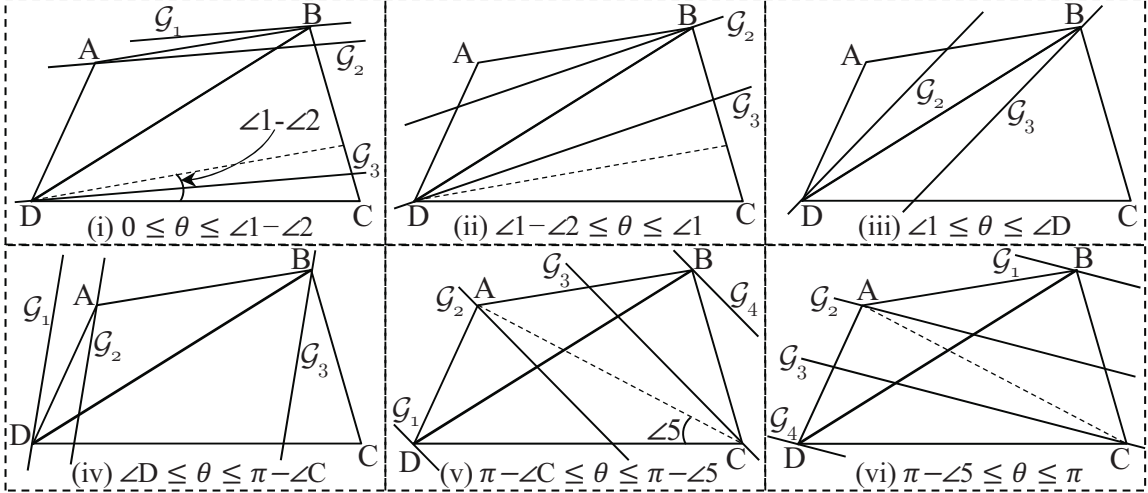}
\vspace{-.em}
\caption{KM over the convex quadrangle shown in \figurename~\ref{fig:2-triangles-convex} ($\angle C=\angle BCD, \angle D=\angle ADC$).}
\label{fig:2-triangles-convex-theta}
\vspace{-2.em}
\end{figure}%
\subsection{Two Triangles Share a Side, Forming a Convex Quadrangle} $\square ABCD$ is a convex quadrangle formed by two arbitrary triangles, $\triangle ABD$ with $|AB|=a$ and $|AD|=d$, and $\triangle BCD$ with $|CB|=b$ and $|CD|=c$, as shown in \figurename~\ref{fig:2-triangles-convex}. $|BD|=e$, $|AC|=f$, $||\triangle ABD||=S_1$, and $||\triangle BCD||=S_2$. WLOG, let $DC$ be on the $x$-axis. The relationship between $\angle 1$ and $\angle 2$ and that between $\angle 3$ and $\angle 4$ determine the shape of the quadrangle, leading to the following four cases:
 \begin{align*}
  {\rm [a]}:&
   \begin{array}{l}
     \angle 1\geq\angle 2,~\angle 3\geq\angle 4,
   \end{array}
  &{\rm [b]}:
  	\begin{array}{r}
  	  \angle 1\geq\angle 2,~\angle 3\leq\angle 4,
  	\end{array}\\
  {\rm [c]}:&
  	\begin{array}{l}
  	  \angle 1\leq\angle 2,~\angle 3\geq\angle 4,
  	\end{array}
  &{\rm [d]}:
  	\begin{array}{r}
  	  \angle 1\leq\angle 2,~\angle 3\leq\angle 4.
  	\end{array}
 \end{align*}
Case [a] is the same as [d] if $BA$ is on $x$-axis after rotating the quadrangle. Similarly, case [b] and [c] are essentially the same. Different cases correspond to different ranges of the orientation angle $\theta$. Below, we will use case [a] to show how to obtain the PDD between the two triangles.

\begin{figure}[!t]
\renewcommand{\algorithmicrequire}{\textbf{Input:}}
\renewcommand\algorithmicensure {\textbf{Output:}}
 \rule[-20pt]{1\textwidth}{0.12em}
 \begin{algorithmic}[1]
 \REQUIRE
Parameters with regard to the convex quadrangle in \figurename~\ref{fig:2-triangles-convex}. 
 \ENSURE
 PDF: $f_D(d)$.
 \STATE $f_D(d)=0$;
 \FOR{($\theta=0$; $\theta\leq\pi$; $\theta=\theta+\delta\theta$)}
	\IF{$\theta\in\mathrm{(i)}~\OR~\mathrm{(iv)}$}
		\STATE $p_m=p_1+p_2$;
	\ELSIF{$\theta\in\mathrm{(ii)}~\OR~\mathrm{(iii)}$}
		\STATE $p_m=p_2$;
	\ELSIF{$\theta\in\mathrm{(v)}~\OR~\mathrm{(vi)}$}
		\STATE $p_m=p_1+p_2+p_3$;
	\ENDIF
    \FOR{($p=0$; $p\leq p_m$; $p=p+\delta p$)}
		\STATE Calculate $l_1$ and $l_3$ for the line $\mathcal{G}$($\theta,p$) according to trigonometry;
		\STATE $l_{low}=l_1$; $l_{up}=l_3$; /* $l_1\leq l_3$ */
		\IF{$l_{low}>l_{up}$}
			\STATE $l_{low}=l_3$; $l_{up}=l_1$; /* $l_3<l_1$ */
		\ENDIF
		\IF{$0\leq d\leq l_{low}$}
			\STATE $f_D(d)=f_D(d)+\frac{2\cdot d^2}{S_1S_2}$;
		\ELSIF{$d\leq l_{up}$}
			\STATE $f_D(d)=f_D(d)+\frac{2\cdot d\cdot l_{low}}{S_1S_2}$;
		\ELSIF{$d\leq l_1+l_3$}
			\STATE $f_D(d)=f_D(d)+\frac{2\cdot d\cdot(l_1+l_3-d)}{S_1S_2}$;
		\ENDIF
    \ENDFOR
 \ENDFOR
 \end{algorithmic}
 \vspace{-1em}
 \rule[0pt]{1\textwidth}{0.05em}
  \vspace{-1em}
 \caption{{Algorithmic procedure for the PDD between two triangles sharing a common side and forming a convex quadrangle}.}\label{alg:2-triangle-convex}
 \vspace{-2.em}
\end{figure}
\begin{figure}[!t]
\centering
\includegraphics[width=0.6\textwidth]{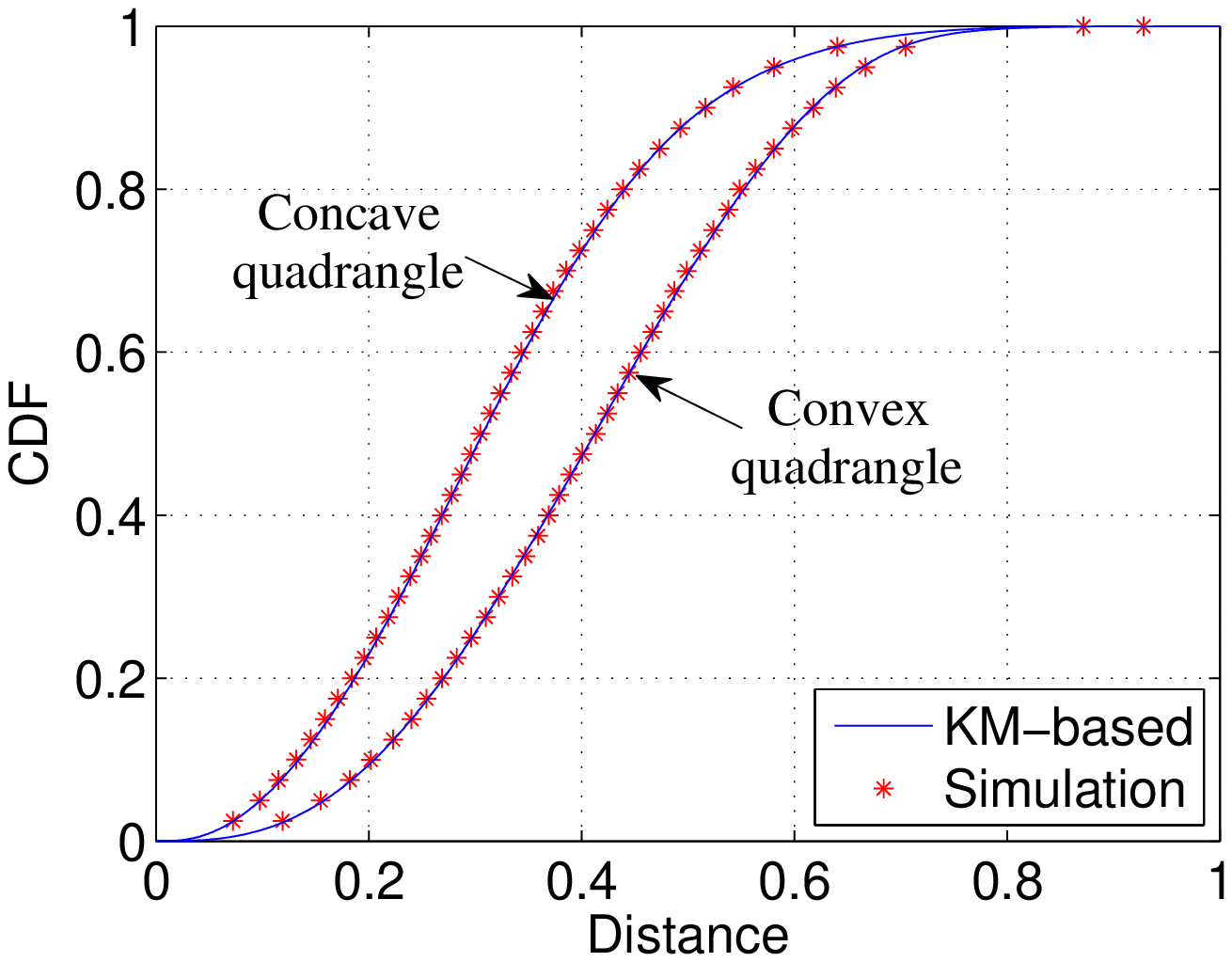}
\vspace{-.em}
\caption{PDD between two triangles sharing a common side and forming a convex/concave quadrangle.}\label{fig:pdd-2-triangles-convex-concave}
\vspace{-.5em}
\end{figure}%
As shown in \figurename~\ref{fig:2-triangles-convex-theta}, for the convenience of calculation based on trigonometry, there are six subcases for case [a] in terms of the range of the line orientation $\theta$ with regard to $x$-axis. For a given $\theta$, only the lines intersecting with both triangles are considered, i.e., the parallel lines between $\mathcal{G}_1$ and $\mathcal{G}_3$ for subcase (i) and (iv), those between $\mathcal{G}_2$ and $\mathcal{G}_3$ for (ii) and (iii), and those between $\mathcal{G}_1$ and $\mathcal{G}_4$ for (v) and (vi). For each line, denote the length of the segment in $\triangle ABD$ as $l_1$, and that in $\triangle BCD$ as $l_3$ ($l_2=0$ in this case). The distances between $\mathcal{G}_1$ and $\mathcal{G}_2$, $\mathcal{G}_2$ and $\mathcal{G}_3$, and $\mathcal{G}_3$ and $\mathcal{G}_4$ are $p_1$, $p_2$, and $p_3$, respectively. The complete derivation process is shown in \figurename~\ref{alg:2-triangle-convex}. Figure \ref{fig:pdd-2-triangles-convex-concave} shows a close match with the simulation results, given the two triangles, $(\angle A=\frac{120\pi}{180},\angle4=\frac{35\pi}{180},\angle 2=\frac{25\pi}{180})$ and $(\angle C=\frac{80\pi}{180}, \angle1=\angle3=\frac{50\pi}{180})$ as an example, as shown in \figurename~\ref{fig:2-triangles-convex}.

\subsection{Two Triangles Share a Side, Forming a Concave Quadrangle} 
\begin{figure}[!t]
\centering
\includegraphics[width=6in]{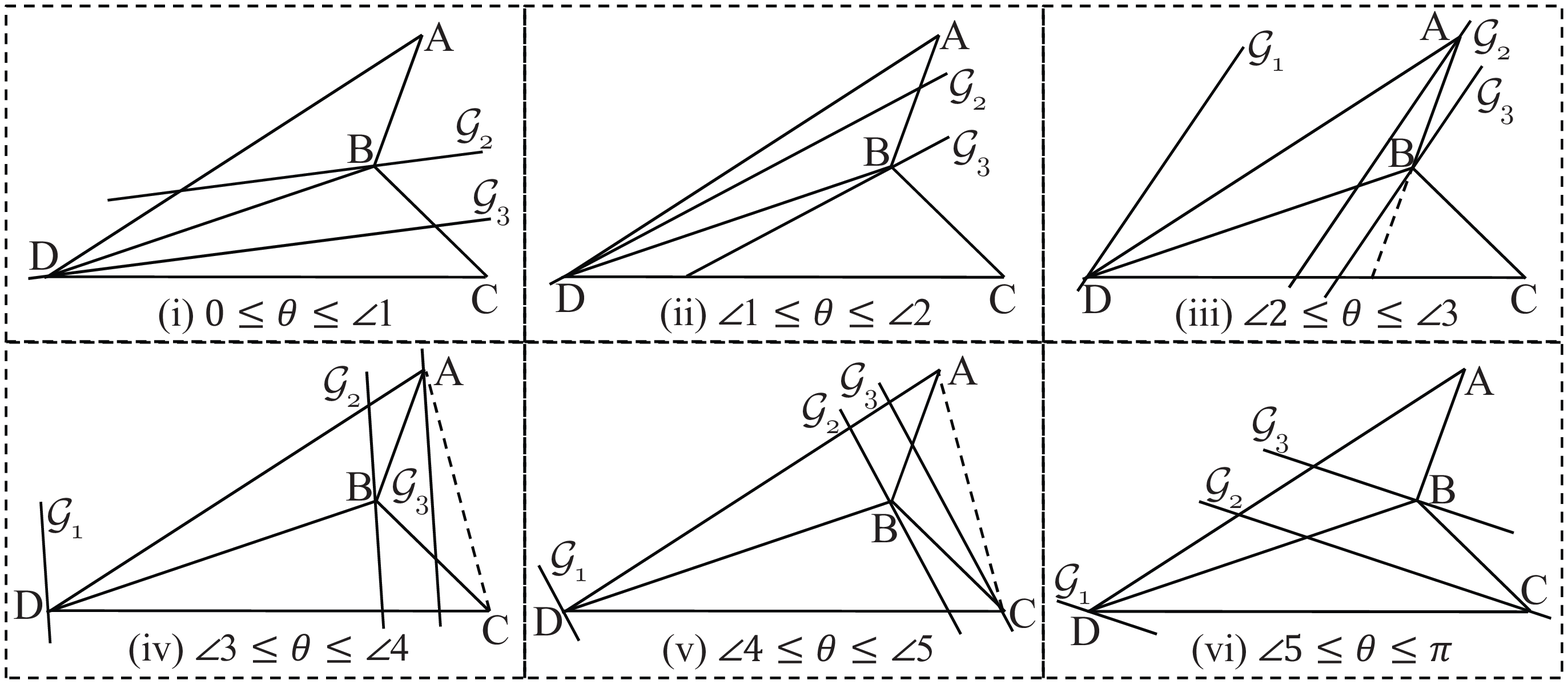}
\vspace{-.em}
\caption{KM over the concave quadrangle shown in \figurename~\ref{fig:2-triangles-concave}.}
\label{fig:2-triangles-concave-theta}
\vspace{-2.em}
\end{figure}%

\begin{figure}[!t]
\renewcommand{\algorithmicrequire}{\textbf{Input:}}
\renewcommand\algorithmicensure {\textbf{Output:}}
 \rule[-20pt]{1\textwidth}{0.12em}
 \begin{algorithmic}[1]
 \REQUIRE
Parameters with regard to the concave quadrangle in \figurename~\ref{fig:2-triangles-concave}.
 \ENSURE
 PDF: $f_D(d)$.
 \STATE $f_D(d)=0$;
 \FOR{($\theta=0$; $\theta\leq\pi$; $\theta=\theta+\delta\theta$)}
	\IF{$\theta\in\mathrm{(i)}~\OR~\mathrm{(ii)}$}\label{alg3:1}
		\STATE $p_m=p_2$;
	\ELSIF{$\theta\in\mathrm{(iii)},\mathrm{(iv)},\mathrm{(v)}~\OR~\mathrm{(vi)}$}
		\STATE $p_m=p_1+p_2$;
	\ENDIF\label{alg3:2}
    \FOR{($p=0$; $p\leq p_m$; $p=p+\delta p$)}
		\STATE Calculate $l_1$, $l_2$, and $l_3$ for the line $\mathcal{G}$($\theta,p$) according to trigonometry;\label{alg3:3}
		\STATE $l_{low}=l_1+l_2$; $l_{up}=l_2+l_3$; $l_s=l_1$; /* $l_1<l_3$ */
		\IF{$l_{low}>l_{up}$}
			\STATE $l_{low}=l_2+l_3$; $l_{up}=l_1+l_2$; $l_s=l_3$; /* $l_3<l_1$ */
		\ENDIF
		\IF{$l_2\leq d\leq l_{low}$}
			\STATE $f_D(d)=f_D(d)+\frac{2d\cdot(d-l_2)}{S_1S_2}$;
		\ELSIF{$d\leq l_{up}$}
			\STATE $f_D(d)=f_D(d)+\frac{2\cdot d\cdot l_{s}}{S_1S_2}$;
		\ELSIF{$d\leq l_1+l_2+l_3$}
			\STATE $f_D(d)=f_D(d)+\frac{2\cdot d\cdot(l_1+l_2+l_3-d)}{S_1S_2}$;
		\ENDIF
    \ENDFOR
 \ENDFOR
 \end{algorithmic}
 \vspace{-1em}
 \rule[0pt]{1\textwidth}{0.05em}
  \vspace{-1em}
 \caption{{Algorithmic procedure for the PDD between two triangles sharing a common side and forming a concave quadrangle.}}\label{alg:2-triangle-concave}
 \vspace{-2.em}
\end{figure}
A concave quadrangle formed by two triangles is shown in \figurename~\ref{fig:2-triangles-concave}. Similarly, for the convenience of calculation by using trigonometry, the line orientation $\theta$ with regard to $x$-axis can be categorized into six cases, as shown in \figurename~\ref{fig:2-triangles-concave-theta}. Given $\theta$, for each line intersecting with both triangles, the length of the segment in $\triangle ABD$ is $l_1$ and that in $\triangle BCD$ is $l_3$. The length of the segment between the above two is $l_2$. Note that $l_2=0$ when the line intersects with $DB$. The distances between $\mathcal{G}_1$ and $\mathcal{G}_2$, and $\mathcal{G}_2$ and $\mathcal{G}_3$ are $p_1$ and $p_2$, respectively. Figure~\ref{alg:2-triangle-concave} summarizes the derivation process. For verification, the following two triangles are used: $(\angle ABD=\frac{110\pi}{180},\angle DAB=\frac{40\pi}{180},\angle ADB=\frac{30\pi}{180})$ and  $(\angle CBD=\frac{160\pi}{180},\angle BCD=\frac{15\pi}{180},\angle CDB=\frac{5\pi}{180})$. The analytical results in close match with the simulation results are shown in \figurename~\ref{fig:pdd-2-triangles-convex-concave}.

 \subsection{Two Triangles Share a Vertex} 
\begin{figure}[!t]
\centering
\subfigure[two triangles in a pentagon: \textit{R1}\&\textit{R3}]{\includegraphics[width=0.333\textwidth]{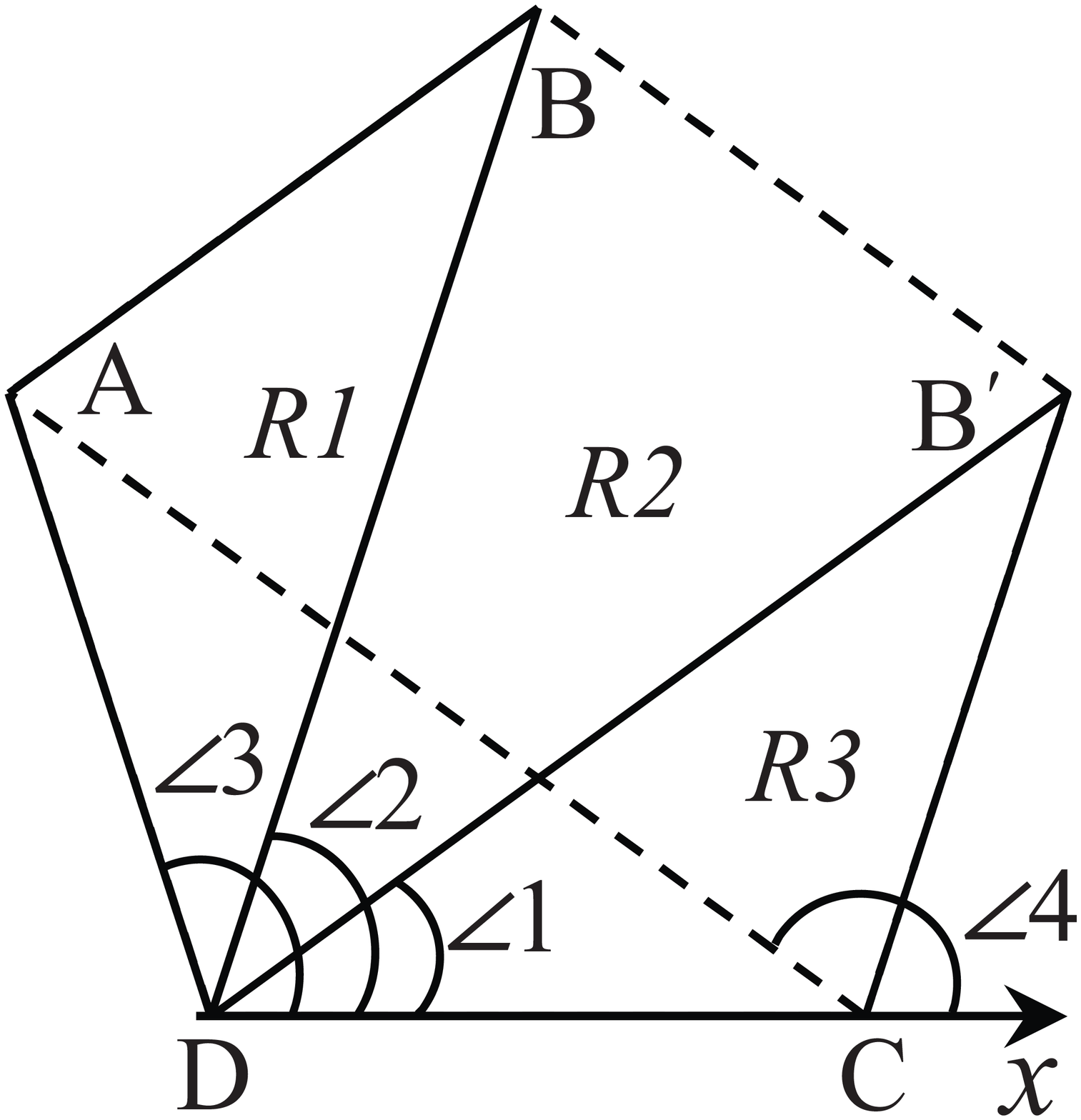}\label{fig:2-triangles-vertex}}
\hspace{2em}
\subfigure[two triangles in a hexagon: \textit{R1}\&\textit{R3}, \textit{R1}\&\textit{R4}, and \textit{R2}\&\textit{R4}]{\includegraphics[width=0.53\textwidth]{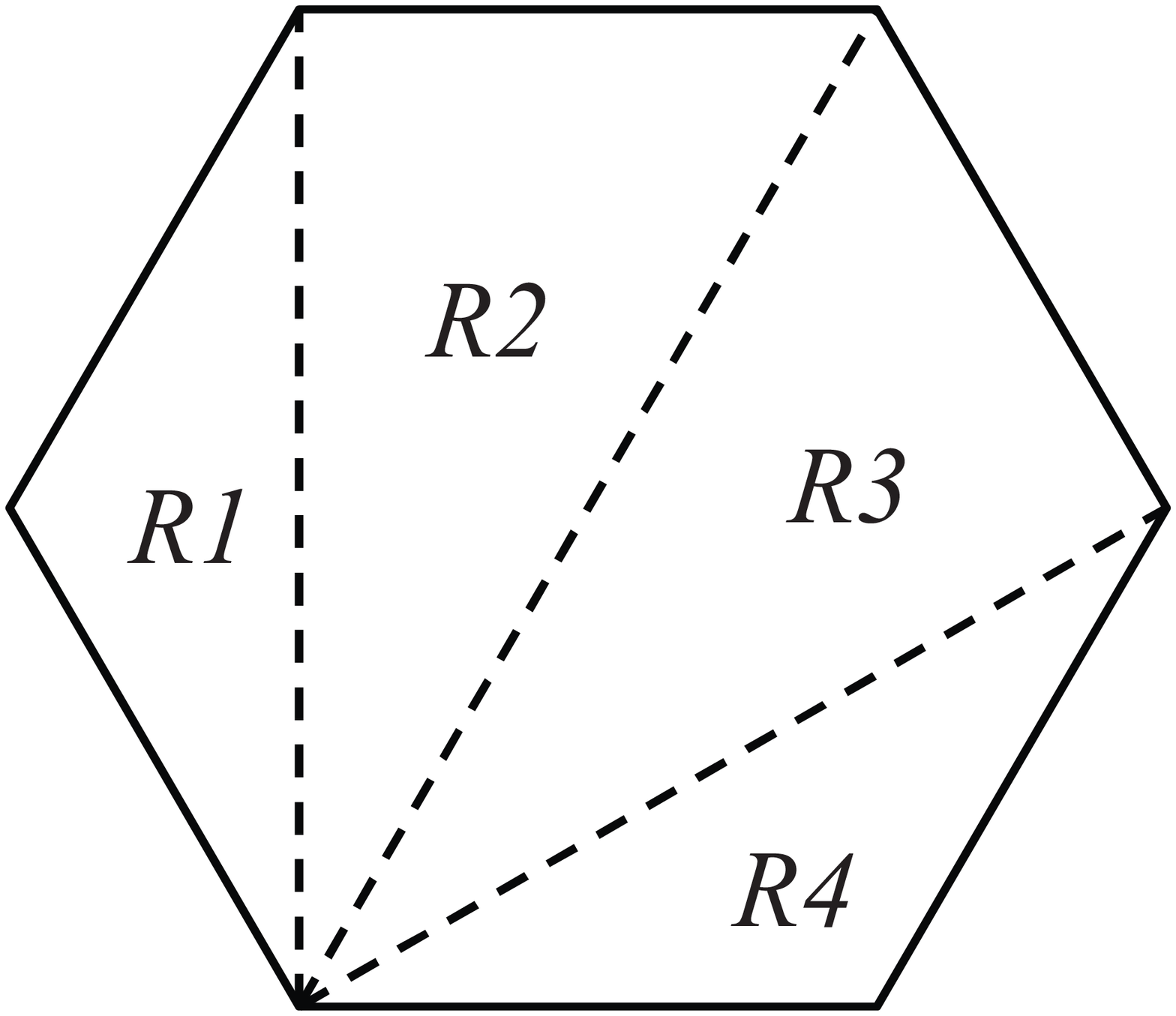}\label{fig:hexagon-2-triangles-vertex}}
\vspace{-.5em}
\caption{Two triangles within a polygon sharing a vertex.}
\vspace{-1.5em}
\end{figure}%
\begin{figure}[!t]
\centering
\subfigure[PDD between two triangles in a pentagon]{\includegraphics[width=0.49\textwidth]{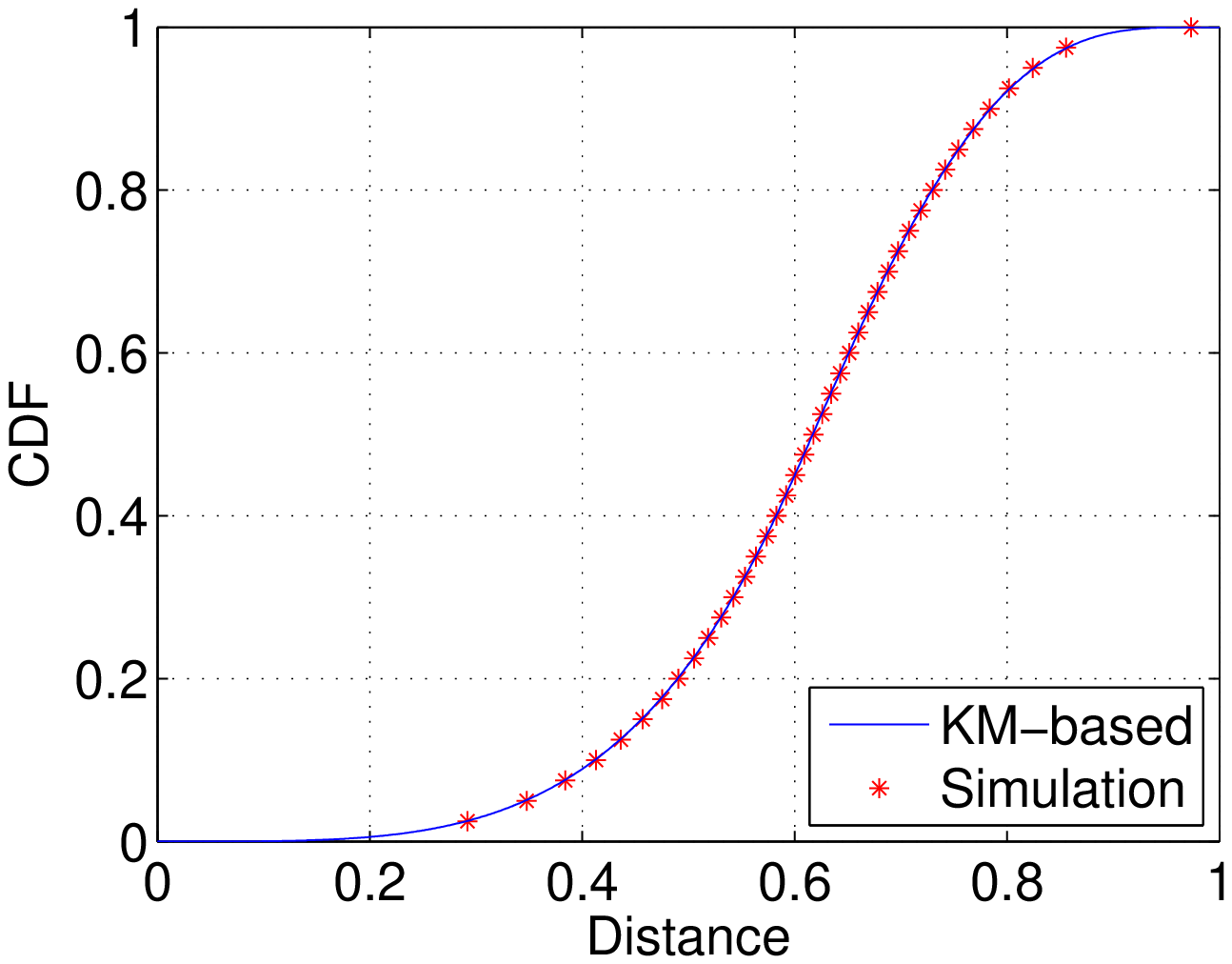}\label{fig:pdd-2-triangles-vertex}}
\subfigure[PDD between two triangles in a hexagon]{\includegraphics[width=0.49\textwidth]{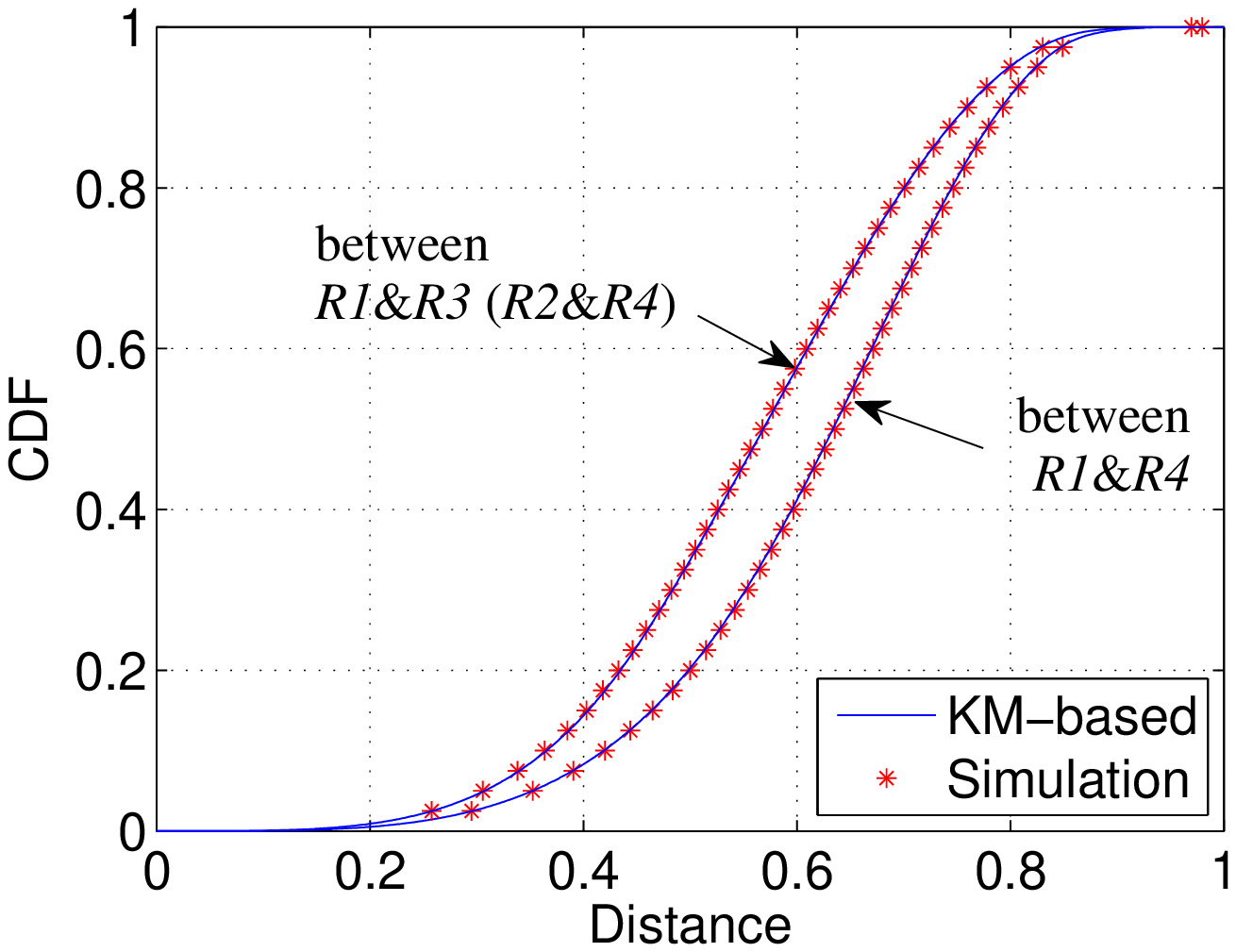}\label{fig:gdd-hexagon-2-triangles-vertex}}
\vspace{-.5em}
\caption{PDD between two triangles in a polygon sharing a vertex.}
\vspace{-2.em}
\end{figure}%
For the simplicity of dividing the range of the line orientation, two special cases are employed for demonstration: [a] two triangles sharing a common vertex are within a regular pentagon, as shown in \figurename~\ref{fig:2-triangles-vertex}, and [b] two triangles sharing a common vertex are within a regular hexagon, as shown in \figurename~\ref{fig:hexagon-2-triangles-vertex}. Note that our approach is not limited to regular polygons, but also applies to other cases associated with arbitrary triangles.

For case [a], the two triangles are labeled by \textit{R1} and \textit{R3}, respectively. With $DC$ on $x$-axis, the line orientation $\theta$ with regard to $x$-axis falls into five subcases: (i) $0\leq\theta\leq\angle1$; (ii) $\angle1\leq\theta\leq\angle2$; (iii) $\angle2\leq\theta\leq\angle3$; (iv) $\angle3\leq\theta\leq\angle4$; (v) $\angle4\leq\theta\leq\pi$. Note that the lines with $\theta$ in (ii) will not be considered, since none of them intersects with both of the triangles. The algorithmic procedure shown in \figurename~\ref{alg:2-triangle-concave} can still obtain the PDD between the two triangles, only with the difference in calculating $p_m$ (line \ref{alg3:1}--\ref{alg3:2}), $l_1$, $l_2$, and $l_3$ (line \ref{alg3:3}). Figure~\ref{fig:pdd-2-triangles-vertex} shows the comparison in a close match with simulation. For case [b] where the regular hexagon is triangulated as shown in \figurename~\ref{fig:hexagon-2-triangles-vertex}, the two triangles are either \textit{R1} and \textit{R3}, \textit{R1} and \textit{R4}, or \textit{R2} and \textit{R4}. According to the symmetry of the regular hexagon, the PDD between \textit{R1} and \textit{R3} is identical with that between \textit{R2} and \textit{R4}. With the same method, the results are shown in \figurename~\ref{fig:gdd-hexagon-2-triangles-vertex}.
\section{PDDs for Arbitrary Polygons}
\begin{figure}[!t]
\centering
\subfigure[PDD within a pentagon]{\includegraphics[width=0.49\textwidth]{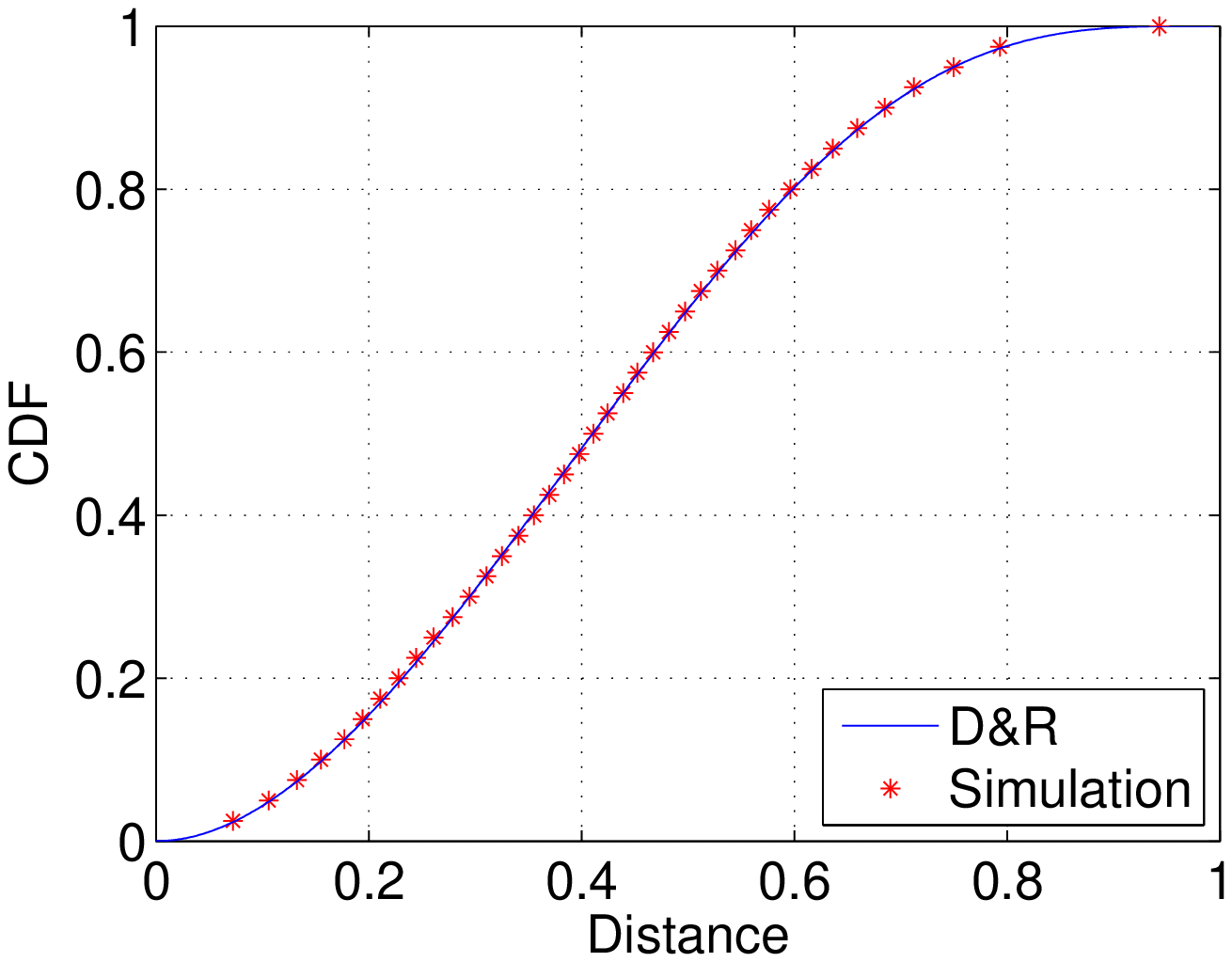}\label{fig:gdd-pentagon}}
\subfigure[PDD within a hexagon]{\includegraphics[width=0.49\textwidth]{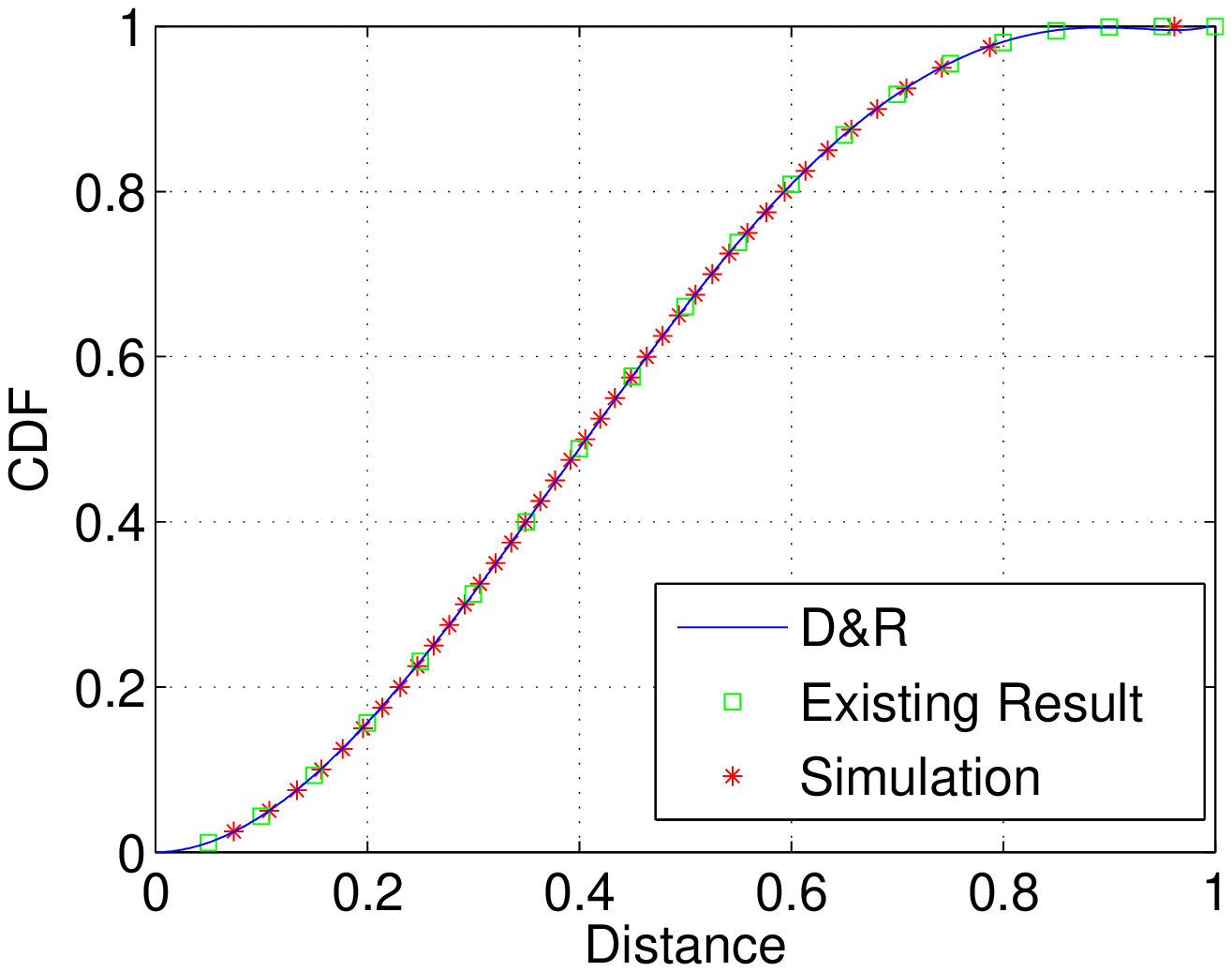}\label{fig:gdd-hexagon}}
\vspace{-.5em}
\caption{PDD within a polygon.}\label{fig:gdd-pentagon-hexagon}
\vspace{-2.em}
\end{figure}
With the triangle-PDDs obtained above, the PDD associated with arbitrary polygons can be obtained through a Decomposition and Recursion (D\&R) approach, since any polygon can be triangulated. Therefore, the PDD-based performance metrics of wireless networks associated with arbitrary polygons can be quantified accurately.

Taking the regular pentagon with the triangulation shown as in \figurename~\ref{fig:2-triangles-vertex} for example, through D\&R, the CDF of the PDD within the pentagon is given by a probabilistic sum,
\begin{equation}
\begin{array}{l}
 F=\frac{2S_\textit{1}}{S}(\frac{S_\textit{1}}{S}F_{\textit{11}}+\frac{S_\textit{3}}{S}F_{\textit{13}}+\frac{S_\textit{2}}{S}F_{\textit{12}})+\frac{S_\textit{3}}{S}(\frac{S_\textit{3}}{S}F_{\textit{33}}+\frac{2S_\textit{1}}{S}F_{\textit{13}})\nonumber~,
\end{array}
\end{equation}
where $S$ is the area of the pentagon. The comparison with simulation is shown in \figurename~\ref{fig:gdd-pentagon}. Likewise, the CDF of the PDD within the regular hexagon of area $S$, with the polygon triangulation shown as in \figurename~\ref{fig:hexagon-2-triangles-vertex}, is
\begin{equation}
\renewcommand{\arraystretch}{1.15}
\begin{array}{l}
 F=\frac{2S_\textit{1}}{S}(\frac{S_\textit{1}}{S}F_{\textit{11}}+\frac{S_\textit{2}}{S}F_{\textit{12}}+\frac{S_\textit{3}}{S}F_{\textit{13}}+\frac{S_\textit{4}}{S}F_{\textit{14}})+\frac{2S_\textit{2}}{S}(\frac{S_\textit{1}}{S}F_{\textit{12}}+\frac{S_\textit{2}}{S}F_{\textit{22}}+\frac{S_\textit{3}}{S}F_{\textit{23}}+\frac{S_\textit{4}}{S}F_{\textit{24}})\nonumber~.
\end{array}
\end{equation}%
The comparison with the existing result obtained in~\cite{hexagons} and simulation in a close match is shown in \figurename~\ref{fig:gdd-hexagon}.

\section{PDDs for Ring Geometries}
\begin{figure}[!t]
\centering
\subfigure[Square ring]{\includegraphics[width=0.4\textwidth]{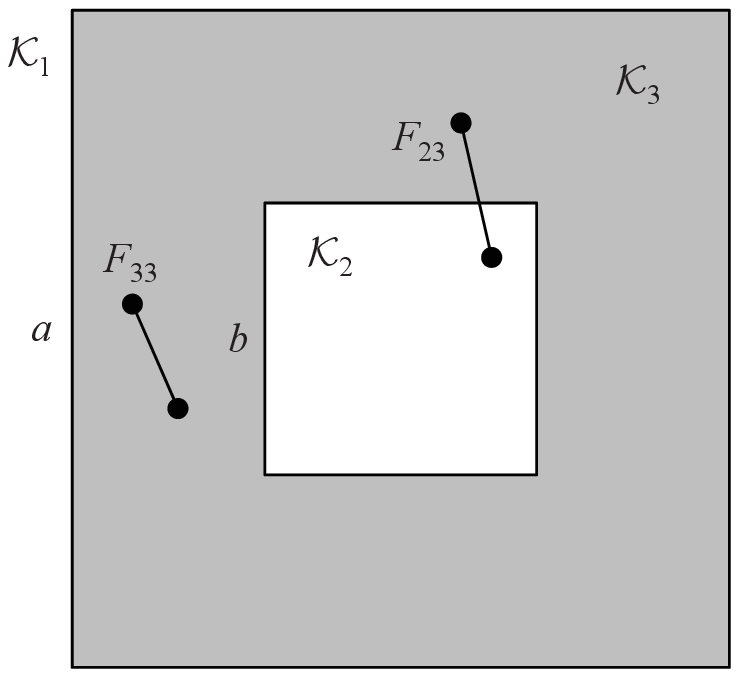}\label{fig:square_ring}}
\hspace{2em}
\subfigure[Hexagon ring]{\includegraphics[width=0.42\textwidth]{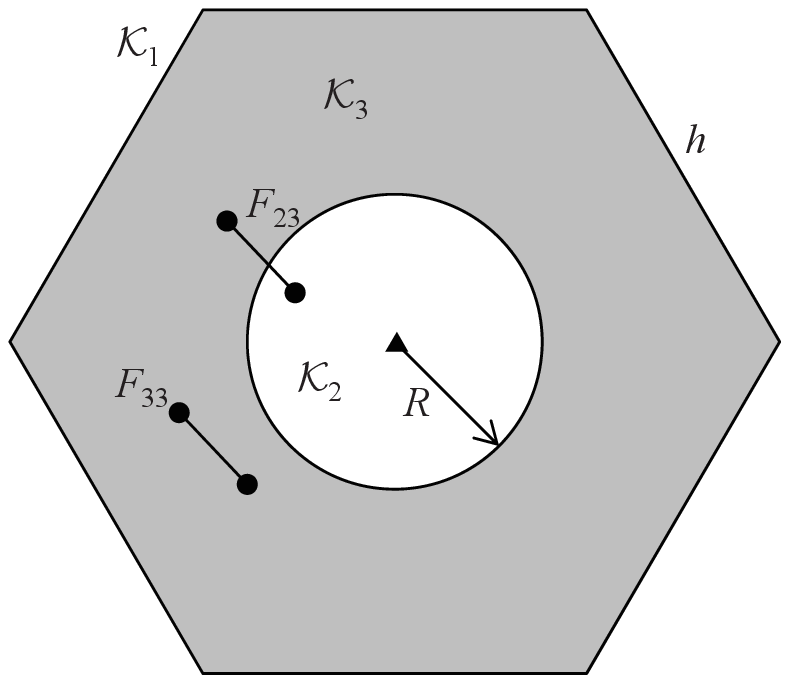}\label{fig:hexagon_ring}}
\vspace{-.5em}
\caption{Two ring geometries.}
\vspace{-1.em}
\end{figure}%
\begin{figure}[!t]
\centering
\subfigure[PDDs associated with square ring]{\includegraphics[width=0.49\textwidth]{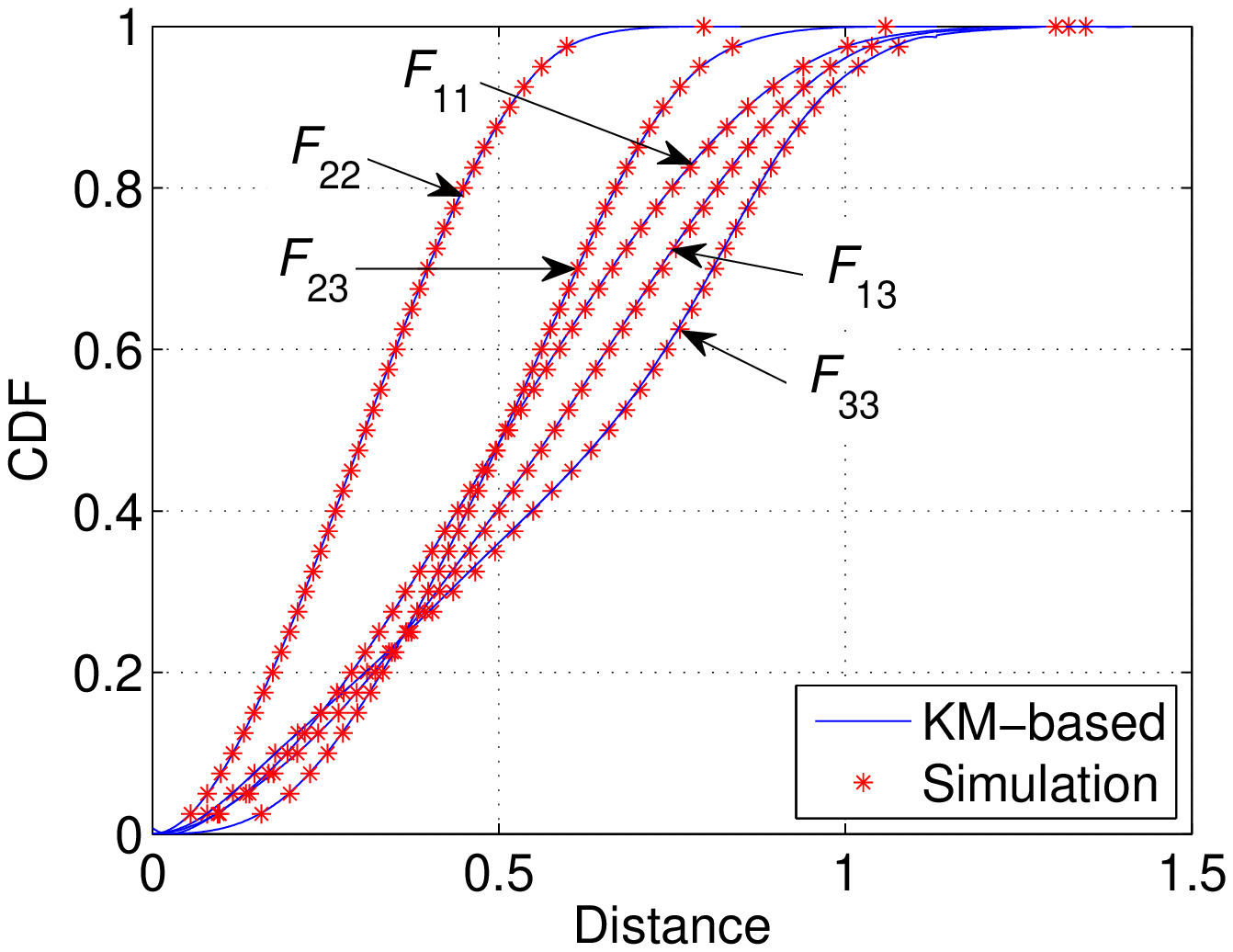}\label{fig:square_ring_cdf}}
\subfigure[PDDs associated with hexagon ring]{\includegraphics[width=0.49\textwidth]{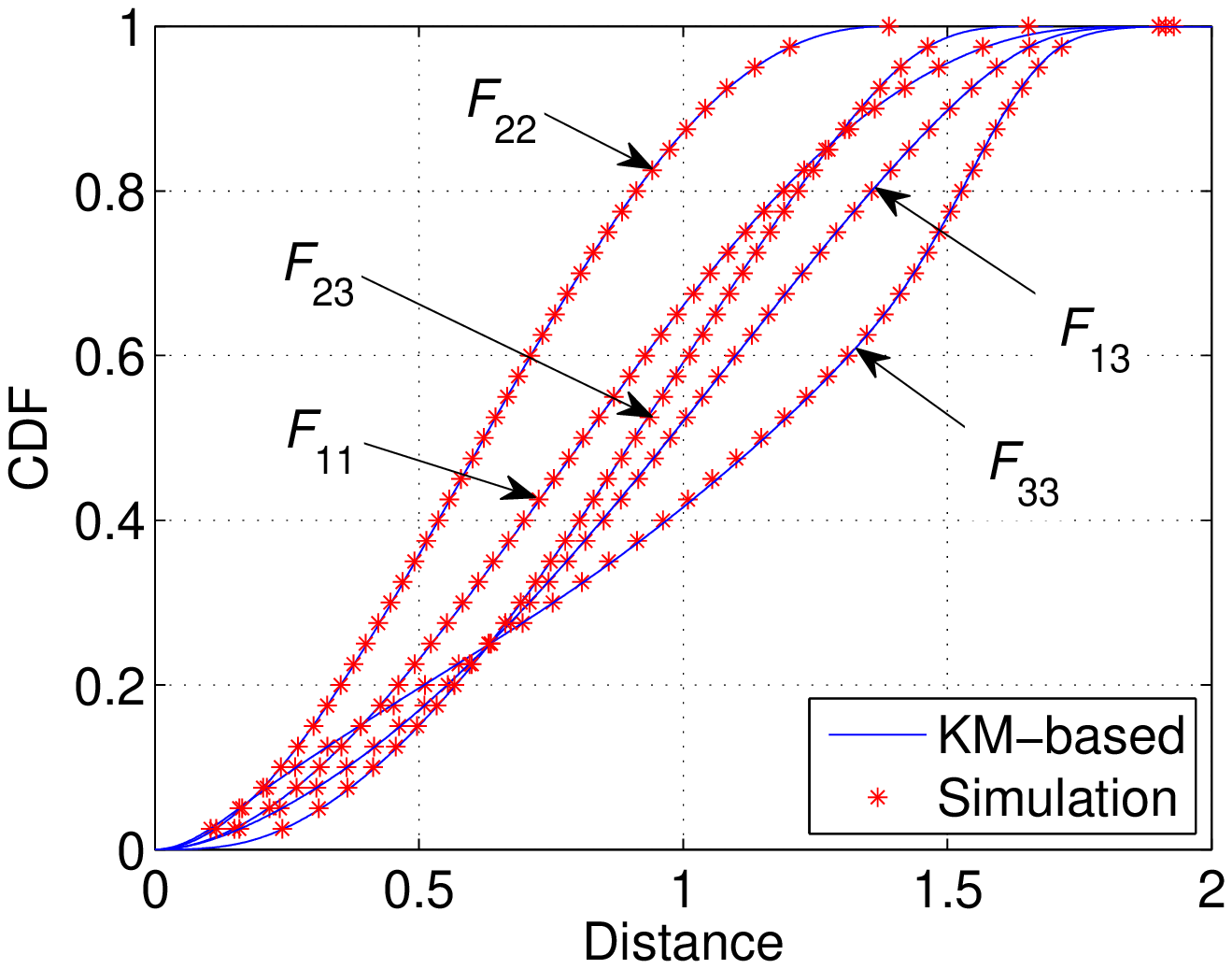}\label{fig:hexagon_ring_cdf}}
\vspace{-.5em}
\caption{PDDs associated with ring geometries.}\label{fig:ring}
\vspace{-2.em}
\end{figure}
In this section, we show the PDDs associated with ring geometries by applying the D\&R approach. For ease of presentation, we use two examples, one is a square ring as shown in \figurename~\ref{fig:square_ring}, and the other is a hexagon ring as shown in \figurename~\ref{fig:hexagon_ring}.

For the square ring, we consider a unit square (i.e., the side length of the square is $a=1$) labeled by $\mathcal{K}_1$, with a smaller square (its side length is $b=0.6$) located in the center and labeled by $\mathcal{K}_2$. The ring area in grey is labeled by $\mathcal{K}_3$. We use $F_{xx}$ to denote the CDF of the random distances within $\mathcal{K}_x$, and $F_{xy}$ the CDF of the random distances between $\mathcal{K}_x$ and $\mathcal{K}_y$. $F_{22}$ and $F_{23}$ can be obtained with the developed approach directly. Then with a weighted probabilistic sum,
\begin{equation}\label{eq:F_holes_prob_sum}
\renewcommand{\arraystretch}{1.15}
\begin{array}{rl}
 F_{\textit{11}}&=\frac{S_\textit{2}}{S_\textit{1}}(\frac{S_\textit{2}}{S_\textit{1}}F_{\textit{22}}+\frac{S_\textit{3}}{S_\textit{1}}F_{\textit{23}})+\frac{S_\textit{3}}{S_\textit{1}}(\frac{S_\textit{2}}{S_\textit{1}}F_{\textit{23}}+\frac{S_\textit{3}}{S_\textit{1}}F_{\textit{33}})=\frac{S_\textit{2}}{S_\textit{1}}F_{\textit{12}}+\frac{S_\textit{3}}{S_\textit{1}}F_{\textit{13}}~,
\end{array}
\end{equation}%
based on which $F_{\textit{12}}$, $F_{\textit{13}}$, and $F_{\textit{33}}$ can be obtained. The obtained CDFs of the PDDs of interest are shown in \figurename~\ref{fig:square_ring_cdf}. Similarly, the CDFs of the PDDs associated with the hexagon ring are shown in \figurename~\ref{fig:hexagon_ring_cdf}, where the radius of the circle in the center of the unit hexagon (its side length is $h=1$) is $R=0.7$. Note that if there are nodes deployed in $\mathcal{K}_2$ and $\mathcal{K}_3$ but with different node densities from each other, the above weighted probabilistic sum is still applicable with different weights due to the node density differences, which shows the way of handling nonuniform node distributions.

\section{Conclusions}
In this report, we first applied the proposed algorithmic approach to obtain the random PDDs associated with arbitrary triangles (triangle-PDDs). Since any polygons can be triangulated, we then used the decomposition and recursion approach for arbitrary polygons based on triangle-PDDs. Finally, the PDDs associated with ring geometries were also shown. The algorithmic procedures were provided to show the mathematical derivation process, based on which either the closed-form expressions or the algorithmic results can be obtained. Together with~\cite{ref2rand}, and the decomposition and recursion approach, all random distances associated with random polygons, regardless between two random points or with an arbitrary reference point, can be obtained, so for any arbitrary geometry shapes with any needed approximation precision.

\section*{Acknowledgment}
This work is supported in part by the NSERC, CFI and BCKDF. We also thank Dr. Lin Cai for her constructive suggestions and encouragement, and Drs. Lei Zheng and Maryam Ahmadi for their comments and suggestions to this work.

\bibliographystyle{abbrv}

\end{document}